\documentstyle[prl,aps,psfig]{revtex}

\begin{document}
\title{\bf Exact Asymptotic Results for Persistence in the Sinai model with Arbitrary Drift}
 
\author{ Satya N. Majumdar$^1$ and Alain Comtet$^2,^3$}

\address{
{\small $^1$Laboratoire de Physique Quantique (UMR C5626 du CNRS), Universit\'e Paul
        Sabatier, 31062 Toulouse Cedex. France}\\
{\small $^2$Laboratoire de Physique Th\'eorique et Mod\`eles Statistiques,
        Universit\'e Paris-Sud. B\^at. 100. 91405 Orsay Cedex. France}\\
{\small $^3$Institut Henri Poincar\'e, 11 rue Pierre et Marie Curie, 75005 Paris, France}}

\date{\today}

\maketitle

\begin{abstract} 
We obtain exact asymptotic results for the disorder averaged persistence
of a Brownian particle moving in a biased Sinai landscape. We employ
a new method that maps the problem of computing the persistence to the problem
of finding the energy spectrum of a single particle quantum Hamiltonian, which
can be subsequently found.  Our method allows
us analytical access to arbitrary values of the drift (bias), thus going beyond the 
previous methods which provide results only in the limit of vanishing drift.
We show that on varying the drift the persistence displays a variety of rich asymptotic behaviors
including, in particular, interesting qualitative changes at some special
values of the drift.

\noindent

\end{abstract}

\section{Introduction}

Persistence, i.e., the probability that a fluctuating field does not change sign upto
time $t$ has been widely studied in recent years in the context of nonequilibrium 
systems\cite{Review}. A wide variety of results, both theoretical and experimental,
are available for pure systems. In contrast, there have been very few studies of persistence
in disordered systems. A notable exception is the study of persistence, 
theoretical\cite{NS} as well as numerical\cite{Jain},
in disordered Ising models. In this paper we study analytically the persistence
in another disordered system namely the celebrated Sinai model\cite{Sinai}, but
in the presence of an additional arbitrary drift. We show that as one varies the drift parameter, 
the disorder averaged persistence displays a wide variety of rich behaviors which
undergo qualitative changes at certain special values of the drift.

The Sinai model\cite{Sinai} is perhaps one of the simplest models of disordered systems
where various disorder averaged physical quantities exhibit rich and
nontrivial behaviors and yet, can be computed analytically\cite{BG}.
Thus the Sinai model serves the role of `Ising' model in disordered systems. 
In this model a Brownian particle undergoes
diffusion in presence of a random time-independent potential. The position
$x(t)$ of the particle evolves via the Langevin equation,
\begin{equation}
{ {dx}\over {dt}}= - {{dU}\over {dx}} + \eta(t),
\label{Langevin}
\end{equation}
where $\eta(t)$ is the thermal noise with $\langle \eta(t)\rangle =0$ and $\langle 
\eta(t)\eta(t')\rangle =\delta(t-t')$ and $U(x)$ is the external potential.
In the biased Sinai model one considers the potential to be simply
$U(x)=-\mu x + \sqrt{\sigma} B(x)$ where  
$B(x)$ represents the trajectory of a Brownian motion in space, i.e., $B(x)=\int_0^x 
\xi(x')dx'$ with $\langle \xi(x)\rangle =0$ and $\langle \xi(x)\xi(x')\rangle =\delta(x-x')$.
The parameter $\mu$ represents the bias or the drift and $\sigma$ measures
the strength of the disorder. Thus the particle 
is subjected, in addition to the thermal noise $\eta(t)$, an external 
position dependent random
force $F(x)= -dU/dx = \mu + \xi(x)$. 
Various physical quantities in the Sinai model have been studied before\cite{BG}.
Here our aim is to compute the persistence
$P(x_0,t)$ defined as the probability that the particle
does not cross the origin upto time $t$ starting at the initial position $x_0\ge 0$ at $t=0$.
Evidently this quantity will depend on the realization of the underlying disorder
potential and will, in general,  vary from one sample of disorder to another. Our final goal
is to compute the disorder averaged persistence $\overline {P(x_0,t)}$ as
a function of both $x_0$ and $t$ for different values of the drift $\mu$.

Physicists have recently studied the persistence in the Sinai
model using various methods
which include an exact probabilistic approach suited for
unbiased Sinai model\cite{CD}, a study
of an equivalent lattice model with random hopping rates\cite{IR}
and also by employing a real space renormalization group method\cite{LMF}.
All of these methods provide asymptotically exact results, but {\it only in the limit of
vanishing drift}, i.e., when $\mu\to 0$. 
Unfortunately, extension of these existing physical methods 
to extract explicit asymptotic 
results for arbitrary nonzero $\mu$ seems rather difficult. Mathematicians,
on the other hand, have studied some aspects of a related quantity namely the distribution of
the first-passage time in the Sinai model for nonzero drift and
some rigorous results seem to exist\cite{KT1,KT2,Tan,HSY}. However, these
mathematical methods are extremely technical and difficult to follow. What is
lacking, so far, is a unified physical approach which, besides reproducing
the known results in a simple and transparent way, provides exact asymptotic results for 
all $\mu$ and yet simple and powerful enough to be easily generalizable 
to other problems. The purpose of this paper is to provide such an approach. The heart of our
approach lies in mapping the problem of computing the persistence in the Sinai model with
arbitrary drift to finding the spectrum of a single particle quantum Hamiltonian,
which can subsequently be done exactly.  

Apart from presenting an unified approach valid for arbitrary drift, there are
two 
other physical motivations for this work.
First, it is well known that the Sinai model
displays a range of intereresting anomalous diffusion properties 
as one tunes the drift $\mu$ through certain finite `critical' values\cite{BG}.
It is therefore theoretically interesting to know how the persistence 
behaviour changes as the drift is varied through these `critical' values.
Secondly and perhaps more importantly, the Sinai model with a nonzero drift
has numerous physical applications\cite{BG} including the diffusion of electrons 
in disordered medium in presence of an electric field, glassy dynamics of 
dislocations in solids, dynamics of random field magnets, dynamics near
the helix-coil transitions in heteroploymers. The most recent application
of the biased Sinai model has been to understand the dynamics of denaturation 
of a single DNA molecule under an external force\cite{LN}. Persistence 
seems to be a natural quantity to study in these systems and hence we
expect that the analytical results presented in this paper will be useful
in many of the physical situations mentioned above.

The paper is organized as follows. In section II, we present our general approach.
A detailed discussion of the pure case ($\sigma=0$) with nonzero drift
is presented in subsection II-A which will help us anticipate the general 
features of persistence in the disordered case studied later in subsection II-B
where we illustrate
the mapping to a quantum mechanics problem. In section III, we discuss the
results for the disorder averaged persistence for positive drift ($\mu>0$).
The results for the negative drift ($\mu<0$), fundamentally different from the
positive drift case, are detailed in  section IV. We conclude in Section V
with a summary and outlook. The details of the derivation of the eigenvalue spectrum of
the quantum Hamiltonian are presented in the appendix-A. In appendix-B, we present
an alternative derivation of the disordered averaged persistence in the case of
positive drift. The details of the second order perturbation theory for negative drift
are presented in appendix-C.     

\section{General Approach}

Consider the particle whose position $x(t)$ evolves via the Langevin equation (\ref{Langevin})
starting initially at $x(t=0)=x_0$. The persistence $P(x_0,t)$ is the
probability that the particle does not cross the origin upto time $t$ starting
at $x_0$. It is also useful to define the distribution of the first-passage time   
$F(x_0,t_*)$ which is simply the probability that the particle hits the origin
for the first time at $t=t_*$ starting initially at $x_0$. The distribution
$F(x_0,t_*)$ is related to the persistence $P(x_0,t)$ via the simple
relation, $P(x_0,t)=1-\int_0^t F(x_0,t_*)dt_*$. This follows from the fact that
the integral $\int_0^t F(x_0,t_*)dt_*$ sums up the probabilities of all the events
when the particle hits the origin before time $t$ and when subtracted from $1$ (which
is the total probability), the resulting quantity, by definition, is the persistence.
Our objective is to first compute $F(x_0,t_*)$ or its Laplace transform and then use
the above relationship to compute the persistence $P(x_0,t)$. 

In order to calculate the first-passage time distribution $F(x_0,t_*)$ we employ
a powerful backward Fokker-Planck approach which has been used before 
to study the persistence in the unbiased Sinai model\cite{CD} as well
as in other contexts\cite{DM}. 
It is instructive to start with a more general quantity,
$Q_p(x_0)={\langle e^{-p\int_0^{t_*} V[x(t')]dt'}\rangle}_{x_0}$ where
$\langle \rangle$ denotes the thermal average, $V[x(t)]$ is an arbitrary functional
and $t_*$ denotes the first-passage time, i.e., the time at which the particle first
hits the origin starting initially at $x_0$ at $t=0$. Note that if we choose $V(x)=1$, 
then $Q_p(x_0)= {\langle e^{-pt_*}\rangle}_{x_0}=\int_0^{\infty} e^{-pt_*}F(x_0,t_*)dt_*$ is 
simply
the Laplace transform of the first-passage time distribution $F(x_0,t_*)$.
For convenience of notations, we will henceforth denote the
initial position $x_0=x$ and the first-passage time $t_*=t$. 

A differential equation for $Q_p(x)$ 
can be derived by evolving the particle from its initial position $x$ over an 
infinitesimal
time $dt$. This gives $Q_p(x) =\langle (1-pV dt)Q_p(x+ dx)\rangle$ where $dx$ is the 
displacement
of the particle in time $dt$ from its initial position $x$. Using Eq. (\ref{Langevin}) one gets
$dx= F(x)dt + \eta(0) dt$ where $F(x)=-dU/dx = \mu+ \sqrt{\sigma} \xi(x)$ is the random force. Expanding
$Q_p(x+ F(x) dt +\eta(0) dt)$ to order $dt$ and averaging 
over
$\eta(0)$, one arrives at the backward Fokker-Planck equation,
\begin{equation}
{1\over {2}} {{d^2 Q_p}\over {dx^2}} + F(x) {{dQ_p}\over {dx}} -p V(x)Q_p=0.
\label{bfp}
\end{equation}
Since here we are interested only in the first-passage time distribution, we will henceforth set $V(x)=1$.
Note, however, that this method is powerful enough to deal with the statistical properties 
of any arbitrary functional $V(x)$ of the stochastic process. We will also assume, without
any loss of generality, that the initial position $x\ge 0$. For $x\le 0$, one will
obtain the same results by changing the sign of the drift $\mu$.
The equation (\ref{bfp}) is supplemented with the two boundary conditions: (i) $Q_p(x=0)=1$, 
since if the particle starts at $x=0$, obviously its first-passage time $t=0$ and
(ii) $Q_p(x\to \infty)=0$,
since the first-passage time $t\to \infty$ if the particle starts at $x=\infty$.

In the next two subsections we discuss the solutions of the differential
equation (\ref{bfp}) respectively for the pure case ($\sigma=0$)
and the disordered case ($\sigma >0$).

\subsection{Pure case with nonzero drift}

It is instructive to discuss first the pure case with nonzero drift ($\mu \ne 0$) in the absence 
of the random potential ($\sigma=0$). The results for the pure case will
help us anticipate what to expect for the disordered case later. 
Solving Eq. (\ref{bfp}) with $F(x)=\mu$ we get
$Q_p(x)=e^{-\left[\mu+\sqrt{\mu^2+2p}\right]x}$ that 
satisfies
the required boundary conditions. Since $Q_p(x)=\int_0^{\infty}e^{-pt}F(x,t)dt$, we
need to invert the Laplace transform which gives
\begin{equation}
F(x,t)= x(2\pi t^3)^{-1/2} 
e^{-(x+\mu t)^2/{2t}}.
\label{fpt}
\end{equation}
Using $P(x,t)=1-\int_0^{t}F(x,t')dt'$, one then gets
\begin{equation}
P(x,t)= 1- {x\over {\sqrt{2\pi}} }\int_0^t {t'}^{-3/2}e^{-(x+\mu t')^2/{2t'}}dt'.
\label{ppers}
\end{equation}
Let us analyze what happens for large $t$ in the three separate cases (i) $\mu>0$, (ii) $\mu<0$
and (iii) $\mu=0$. 
\vspace{0.2cm}

\noindent (i) For positive bias away from the origin ($\mu>0$), the particle eventually 
escapes to $\infty$ with a nonzero probability and $P(x,t) \to P(x)$ as $t\to \infty$. 
Taking $t\to \infty$ limit in Eq. (\ref{ppers}) one easily obtains this
eventual persistence `profile' $P(x)=1-e^{-2\mu x}$. 
\vspace{0.4cm}

\noindent (ii) In the opposite case ($\mu=-|\mu|<0$), it follows from Eq. (\ref{ppers}) that as
$t\to \infty$, $P(x,t)\approx
\sqrt{2/{\pi}}x |\mu|^{-2}t^{-3/2}e^{-(x-|\mu|t)^2/{2t}}$. In this case, a more useful
information is contained in the asymptotic first-passage distribution $F(x,t)$.  
From the exact expression of $F(x,t)$ in Eq. (\ref{fpt}), we find that in the appropriate scaling limit
$x\to \infty$, $t\to \infty$ but keeping $x/t$ fixed, the first-passage distribution approaches a delta 
function, $F(x,t) \to \delta (t- x/|\mu|)$. Equivalently the Laplace transform, 
$Q_p(x)\to e^{-px/|\mu|}$ which also follows directly from the expression of 
$Q_p(x)=\exp\left[-\left(\sqrt{|\mu|^2+2p}-|\mu|\right)x\right]$ in the correct scaling limit
$x\to 0$, $p\to 0$ but keeping the product $px$ fixed. Thus in this case, at late times, the
particle essentially moves ballistically with velocity $|\mu|$ and crosses
the origin for the first time at $t=x/|\mu|$. 
\vspace{0.4cm}

\noindent (iii) In the unbiased case ($\mu=0$), we recover
from Eq. (\ref{ppers}) the well known exact result\cite{Feller}, $P(x,t) = {\rm 
erf}(x/\sqrt{2t})$ where ${\rm erf}(x)={2\over {\sqrt{\pi}}}\int_0^{x}e^{-u^2}du$ is
the error function. 
\vspace{0.2cm}

In the next subsection, we switch on the quenched disorder ($\sigma>0$) and examine the consequences
on the asymptotic properties of the disorder averaged persistence .  

\subsection{Disordered Case with nonzero drift}

Unlike the pure case, we can no longer solve the differential equation (\ref{bfp}) exactly
for the disordered case since $F(x)=\mu +\sqrt{\sigma}\xi (x)$ now has an $x$-dependent
random part. To make further progress we first make a Hopf-Cole transformation, 
$Q_p(x)=\exp\left[-\int_0^{x}z_p(x')dx'\right]$. Clearly $z_p(x)=-d{\log Q_p(x)}/dx$
is a slope variable. By construction, $Q_p(x)$
automatically satisfies the boundary condition $Q_p(x=0)=1$. Substituting
this form of $Q_p(x)$ in Eq. (\ref{bfp}), we find that the slope variable 
$z_p(x)$ satisfies a first order stochastic Riccati equation
\begin{equation}
{{dz_p(x')}\over {dx'}}=z_p^2(x')-2\left[\mu+\sqrt{\sigma}\xi(x')\right] z_p(x')-2p.
\label{riccati}
\end{equation}  
The right hand side of the above equation contains a multiplicative noise term and
we will interpret it according to the Stratonovich prescription. 
Note that since Eq. (\ref{riccati}) is a first order equation, $z_p(x')$ at an arbitrary $x'$
will be fully determined as a functional of the noise history $\{\xi(x')\}$, at least in principle, 
provided the value of $z_p$ is known at some `initial' point. Note that this `initial' point
can be anywhere. The program would 
then be to substitute this fully determined functional to evaluate the integral $\int_0^x z_p(x')dx'$
and thereby determine $Q_p(x)=\exp\left[-\int_0^{x}z_p(x')dx'\right]$. Subsequently one would
perform the disorder average ${\overline {Q_p(x)}}$ where the overbar indicates an
average over the noise history $\{\xi(x')\}$ for fixed $x$.     

However, there is one problem in implementing this program namely the `initial' value of
$z_p(x')$ is not specified. Consequently the solution of the first order equation (\ref{riccati}) 
will involve an unknown parameter, i.e., the `initial' value of $z_p(x')$. There exists, however,
a rather nice trick to get around this difficulty. This trick has been used 
before in the Sinai model in various contexts\cite{BG,CD,MC}. It is useful
to outline this trick in the present context. To use this trick, we first fix $x$ in
the definition $Q_p(x)=\exp\left[-w(x)\right]$ where we have defined $w(x)=\int_0^{x}z_p(x')dx'$.
Keeping $x$ fixed we then make a change of variable, $\tau= x-x'$. Thus, when $x'\to \infty$, 
$\tau\to -\infty$ and when $x'\to 0$, $\tau\to x$. Besides, $x'=x$ corresponds to $\tau=0$ (see Fig. 
1).
Then $w(x)=\int_0^x z_p(x')dx'=-\int_x^0 z_p(x-\tau)d\tau=\int_0^{x}{\tilde z_p}(\tau)d\tau$
where ${\tilde z_p}(\tau)=z_p(x-\tau)$. In this new $\tau$ variable, the equation (\ref{riccati})
becomes,
\begin{equation}
{{d{\tilde z_p}(\tau)}\over {d\tau}}=-{\tilde z_p}^2(\tau)+2\left[\mu+\sqrt{\sigma}{\tilde \xi}(\tau)\right] 
{\tilde z_p}(\tau)+2p ,
\label{riccati1}
\end{equation}
where ${\tilde \xi(\tau)}= \xi(x-\tau)$ and $\tau \in [-\infty,x]$. Note that
$\langle {\tilde \xi}(\tau)\rangle=0$ and $\langle {\tilde \xi}(\tau){\tilde \xi}(\tau')=\delta(\tau-\tau')$.
To simplify further, we 
substitute ${\tilde 
z_p}(\tau)=\exp\left[\phi(\tau)\right]$
in Eq. (\ref{riccati1}) and find that the variable $\phi(\tau)$ satisfies
a much simplified stochastic equation containing only additive noise (and no multiplicative noise)
\begin{equation}
{{d\phi}\over {d\tau}}= b(\phi)+2\sqrt{\sigma}{\tilde \xi}(\tau),
\label{phieq}
\end{equation}
where the source term $b(\phi)$ is given by
\begin{equation}
b(\phi)= -e^{\phi} + 2\mu + 2p\, e^{-\phi}.
\label{bphi}
\end{equation}

What did we gain in the change of variable from $x'\in [0,\infty]$ to $\tau\in [-\infty,x]$?
The point is that we have, via this change of variable, gotten around the problem
of `initialization' of the original variable $z_p(x')$ at any `initial' point. To see
this, let us consider the Eq. (\ref{phieq}) which is valid in the regime $\tau\in [-\infty,x]$.
We can interpret this equation now as a simple Langevin equation describing
the evolution of the position $\phi(\tau)$ of a classical particle with `time'
$\tau$ starting from $\tau=-\infty$. Of course, we still do not know the value
of $\phi(\tau)$ at $\tau=-\infty$. The point, however, is that this initial
condition at $\tau=-\infty$ is completely irrelevant. No matter what this 
initial condition at $\tau=-\infty$ is,
it is clear from Eq. (\ref{phieq}) that eventually when $\tau$ is far away
from its starting point $\tau=-\infty$, the system will approach a stationary
state. This is because the Eq. (\ref{phieq}) describes the noisy `thermal' motion
of a particle in a classical potential $U_{\rm cl}(\phi)= -\int_0^{\phi} b(u)du=
e^{\phi}-2\mu\phi+2pe^{-\phi}-(2p+1)$. Hence the particle will eventually
reach the equilibrium and the stationary probablity distribution of $\phi$
is simply given by the Gibbs measure, 
\begin{equation}
P_{\rm st}(\phi)=A \exp\left[-{1\over {2\sigma}}U_{\rm cl}(\phi)\right]=
A \exp\left[ {1\over {2\sigma}}\int_0^{\phi}b(u)du
\right]
\label{Gibbs}
\end{equation}
where $b(\phi)$ is given by Eq. (\ref{bphi}) and $A$ is a normalization constant
such that $\int_{-\infty}^{\infty}P_{\rm st}(\phi)d\phi=1$. For later purposes
we also define $P_{\rm st}(\phi)=\psi_0^2(\phi)$ where 
\begin{equation}
\psi_0(\phi)=\sqrt{A} \exp\left[{1\over 
{4\sigma}}\int_0^{\phi}b(u)du\right],
\label{psi00}
\end{equation}
the function $b(\phi)$ is given by Eq. (\ref{bphi}) and $A$ is such that
$\int_{-\infty}^{\infty}\psi_0^2(\phi)d\phi=1$.

So now we know that starting at $\tau=-\infty$ with arbitrary initial condition, by
the time the system reaches $\tau=0$, it has already achieved the stationary
measure. But our task is not yet complete. We now have to evolve the system
via its equation of motion (\ref{phieq}) from $\tau=0$ to $\tau=x$ (knowing
that at $\tau=0$ the distribution of $\phi$ is given by the Gibbs measure in Eq. (\ref{Gibbs}))
and evaluate the disorder average
\begin{equation}
{\overline {Q_p(x)}}=E\left[\exp[-w(x)]\right]=E\left[\exp[-\int_0^{x}e^{\phi(\tau)}d\tau ]\right],
\label{expect}
\end{equation}
where $w(\tau)=\int_0^\tau e^{\phi(\tau')}d\tau'$ as defined earlier and 
$E[..]$ denotes the expectation value of the 
random variable inside the parenthesis. Let us introduce the quantity $R(\phi,\tau)=E_\phi[e^{-\lambda w(\tau)}]$
which denotes the expectation value $e^{-\lambda w}$ at time $\tau$ with $\phi(\tau)=\phi$.
More precisely, if $P_J[w(\tau)=w, \phi(\tau)=\phi, \tau]$ denotes the joint probability
distribution of the variables $w(\tau)$ and $\phi(\tau)$ at time $\tau$, then
$R(\phi,\tau)=\int e^{-\lambda w}P_J[w,\phi,\tau]dw$. 
We have introduced the additional parameter $\lambda$ for later convenience whose value will be eventually
set to $\lambda=1$. Note that if we set $\lambda=0$, $R(\phi,\tau)$ is simply the
probability distribution of $\phi$ at time $\tau$. Thus from now on, we will refer
to the $\lambda=0$ case as the `free' problem. When $\lambda=1$, it is clear 
that,
\begin{equation}
{\overline {Q_p(x)}}=E\left[\exp[-w(x)]\right]=\int_{-\infty}^{\infty} R(\phi,x)d\phi .
\label{rphix}
\end{equation}
The advantage for this small detour in introducing the new quantity $R(\phi, \tau)$ is that 
one can now write down a Fokker-Planck
equation for $R(\phi,\tau)$ in a straightforward manner. In fact, incrementing $\tau$ to 
$\tau+d\tau$ in the definition $R(\phi,\tau)= E_\phi\left[\exp[-\lambda \int_0^{\tau} 
e^{\phi(\tau')}d\tau' ]\right]$
and using the Langevin equation (\ref{phieq}), we find that $R(\phi,\tau)$ satisfies the
following equation,
\begin{equation}
{{\partial R}\over {\partial \tau}}=2\sigma {{\partial^2 R}\over {\partial \phi^2}}-b(\phi)
{{\partial R}\over {\partial \phi}}-\left[b'(\phi)+\lambda e^{\phi}\right]R,
\label{Req}
\end{equation}
where $b'(\phi)=db/d\phi$ with $b(\phi)$ given from Eq. (\ref{bphi}). Note that 
at $\tau=0$, $w(0)=0$ and hence $R(\phi,0)$ is just the probability distribution of $\phi$ 
which is given by the Gibbs measure $R(\phi,0)=P_{\rm st}(\phi)$ in Eq. (\ref{Gibbs}).
Starting with this initial condition at $\tau=0$, we need to evolve the equation (\ref{Req}) upto
$\tau=x$, determine $R(\phi,x)$ and then integrate over $\phi$ in Eq. (\ref{rphix}) to 
finally obtain the desired quantity ${\overline {Q_p(x)}}$. Note that for $\lambda=0$, the
Eq. (\ref{Req}) is simply the ordinary Fokker-Planck equation for the probability distribution
of $\phi$ in the `free' problem.
 
We next substitute $R(\phi,\tau)=\exp\left[{1\over 
{4\sigma}}\int_0^{\phi}b(u)du\right] G(\phi,\tau)$ in Eq. (\ref{Req}) to get rid of the 
first derivative term on the right hand side of Eq. (\ref{Req}) and find the
following evolution equation for the Green's function $G(\phi,\tau)$,
\begin{equation}
{{\partial G}\over {\partial \tau}}=2\sigma {{\partial^2 G}\over {\partial \phi^2}}
-\left[{{a}\over {2}}b^2(\phi) +{1\over {2}}b'(\phi) + \lambda e^{\phi}\right]G,
\label{green}
\end{equation}
where $a=1/{4\sigma}$ and $G(\phi,0)= \exp\left[-{1\over
{4\sigma}}\int_0^{\phi}b(u)du\right]R(\phi,0)=\sqrt{A} \psi_0(\phi)$, using
the Gibbs measure in Eq. (\ref{Gibbs}). To solve Eq. (\ref{green}) we make the standard eigenvalue 
decomposition
\begin{equation}
G(\phi, \tau)= \sum_{E} c_{E} g_{E}(\phi) e^{-4\sigma E \tau},
\label{decom}
\end{equation}
where the eigenfunctions $g_E(\phi)$ satisfy the Schr\"odinger equation,
\begin{equation}
-{1\over {2}}{ {d^2 g_E(\phi)}\over {d\phi^2}}+ \left[{{a^2}\over {2}}b^2(\phi) +{a\over {2}}b'(\phi) + a\lambda 
e^{\phi}\right]g_E(\phi)=Eg_E(\phi),
\label{shrodinger}
\end{equation}
with $b(\phi)$ given by Eq. (\ref{bphi}) and $a=1/{4\sigma}$. The coefficients $c_E$'s in Eq. (\ref{decom}) are 
determined from the 
initial condition, $G(\phi,0)= \sqrt{A} \psi_0(\phi)$. Using orthogonality of eigenfunctions one finds,
\begin{equation}
c_E= \sqrt{A} \int_{-\infty}^{\infty} d\phi g_E^* (\phi) \psi_0(\phi)=\sqrt{A} <g_E|\psi_0> ,
\label{coeff}
\end{equation}
where we have used the standard bra-ket notation of quantum mechanics. Note also that if we consider the `free' 
problem by setting $\lambda=0$, it is easy to verify from Eq. (\ref{shrodinger}) that there is an eigenfunction 
with energy $E=0$ which corresponds to the stationary state of the `free' problem.
This zero energy eigenfunction is given precisely by 
$\psi_0(\phi)$ in Eq. (\ref{psi00}) and the Gibbs measure is just the square of 
this eigenfunction, $P_{\rm st}(\phi)=\psi_0^2(\phi)$. 

Substituting the $c_E$'s from Eq. (\ref{coeff}) in the decomposition equation (\ref{decom})
and setting finally $\tau=x$ we get the following expression of $R(\phi,x)$ in terms of the
eigenfunctions,
\begin{equation}
R(\phi,x)= \psi_0^*(\phi) \sum_E <g_E|\psi_0> g_E(\phi) e^{-4\sigma E x}.
\label{rphix1}
\end{equation}
By integrating $R(\phi, x)$ in Eq. (\ref{rphix1}) over $\phi$, we finally obtain 
the disorder average $\overline {Q_p(x)}$ in a compact form,
\begin{eqnarray}
{\overline {Q_p(x)}}= \int_{-\infty}^{\infty}  R(\phi,x)d\phi &=& \int_{-\infty}^{\infty} d\phi \psi_0^*(\phi) 
g_E(\phi) \sum_E <g_E|\psi_0> e^{-4\sigma E x} \nonumber \\
&=& \sum_E <g_E | \psi_0> <\psi_0|g_E> e^{-4\sigma E x} \nonumber \\
&=& <\psi_0 | e^{-4\sigma {\hat H} x}|\psi_0>,
\label{compact}
\end{eqnarray}
where the quantum Hamiltonian $\hat H$ in the $\phi$ basis is given by,
\begin{equation}
{\hat H}= -{1\over {2}}{{\partial^2}\over {\partial \phi^2}} + \left[{{a^2}\over {2}}b^2(\phi) +{a\over 
{2}}b'(\phi) + a\lambda
e^{\phi}\right]= {\hat H_0} + {\hat H_1}.
\label{hamil}
\end{equation}
Here ${\hat H_0}= -{1\over {2}}{{\partial^2}\over {\partial \phi^2}} + \left[{{a^2}\over {2}}b^2(\phi) +{a\over
{2}}b'(\phi)\right]$ is the Hamiltonian of the `free' problem (corresponding to $\lambda=0$),
${\hat H_1}= \lambda a e^{\phi}$ is like a perturbation Hamiltonian and
$b(\phi)=-e^{\phi}+2\mu+2pe^{-\phi}$.

The exact formula ${\overline {Q_p(x)}}=\int_0^{\infty} dt e^{-pt} {\overline {F(x,t)}}=<\psi_0 | e^{-4\sigma {\hat 
H} x}|\psi_0>$ in Eq. (\ref{compact}) is, in fact, the central result of this paper. This result
tells us that the Laplace transform of the disorder averaged first-passage time 
can, in principle, be fully computed for arbitrary starting position $x$, arbitrary $p$ (and hence 
for arbitrary $t$) and also for any value of the drift $\mu$ provided one can compute all the
eigenvalues and the corresponding eigenfunctions of the Hamiltonian $\hat H$ in Eq. (\ref{hamil}).
In other words, the calculation of the disorder averaged persistence is reduced to finding
the spectrum of the quantum Hamiltonian $\hat H$. In the next two sections we show how
this spectrum can be determined in limiting cases which lead to exact asymptotic results (large $t$ limit)
for the persistence $P(x,t)$ for any arbitrary drift $\mu$.

\section{ Explicit results for postive drift ($\mu>0$)}

In this section we focus on the positive drift $(\mu>0)$ case. We have seen in Section II-A that
for the pure case, the particle eventually escapes to infinity with a nonzero probability
when there is a positive drift $(\mu>0)$ away from the origin. 
This escape probability $P(x)$ is precisely the persistence
$P(x,t)$ in the limit $t\to \infty$. Due to a nontrivial dependence of this probability
$P(x)$ on the initial position $x$, we call $P(x)$ the persistence profile. Note that 
from the relationship, $P(x,t)=1-\int_0^t F(x,t_*)dt_*$ where $F(x,t)$ is the first-passage time distribution,
it follows that the persistence profile is given by $P(x)=1-\int_0^{\infty}F(x,t_*)dt_*=1-Q_0(x)$
where we recall that $Q_p(x)=\int_0^{\infty} e^{-pt} F(x,t)dt$ is just the Laplace transform 
of the first-passage time distribution. 

In the disordered case with positive drift, one would expect a similar behavior namely
for each sample of disorder, the particle will eventually escape to infinity with a
nonzero sample dependent probability $P(x)$. The disorder averaged persistence profile
is then given by ${\overline {P(x)}}=1-{\overline {Q_0(x)}}$. Using this relationship
one can then compute, in the large $t$ limit, the exact
time-independent persistence profile by setting $p=0$
in the general formula for ${\overline {Q_p(x)}}$ in Eq. (\ref{compact}) derived
in Section II-B. For $p=0$, we get from Eq. (\ref{bphi}), $b(\phi)=-e^{\phi}+2\mu$.
Substituting this $b(\phi)$ in Eq. (\ref{hamil}), setting $\lambda=1$ and simplifying, we 
find ${\hat H} = {\hat H_M} + \nu^2/8$ where $\nu=\mu/\sigma$ and ${\hat H_M}$ is a generalized Morse
Hamiltonian given by,
\begin{equation}
{\hat H_M}= -{1\over {2}} {{\partial^2}\over {\partial \phi^2}}+ {a^2\over {2}}e^{2\phi}-
{{a(\nu-1)}\over {2}}e^{\phi}.
\label{morse}
\end{equation}  
It then follows from Eq. (\ref{compact}) that 
\begin{equation}
{\overline {Q_0(x)}}=e^{-\sigma \nu^2 x/2}<\psi_0|e^{-4\sigma {\hat H_M} x}|\psi_0>.
\label{compact1}
\end{equation}
To evaluate the matrix element in Eq. (\ref{compact1}) explicitly we need to know the spectrum of 
the generalized Morse 
Hamiltonian ${\hat H_M}$. Fortunately this spectrum can be fully determined. This calculation
is done in details in appendix-A. Here we just summarize this spectrum and use the results
to compute ${\overline {P(x)}}=1-{\overline {Q_0(x)}}$ explicitly.

The spectrum of ${\hat H_M}$ consists of two parts: a discrete part with negative energies that
correspond to the bound states and a continuous part with positive energies corresponding to
the scattering states (see appendix-A). The nature of the spectrum depends on the parameter
$\nu=\mu/\sigma$. It turns out that there is a critical value $\nu_c=2$ such that for
$\nu>\nu_c$, the spectrum has both the bound states and the scattering states. In contrast,
for $\nu<\nu_c$, there are no bound states and only scattering states exist.
We notice that a similar behaviour was obtained in  the study of transport
properties of the Sinai model\cite{MC}. 
The eigenvalues
and the corresponding eigenfunctions are given as follows.
\vspace{\baselineskip}

\noindent {\bf Bound States:} The bound states are labelled by an interger $n$. The eigenvalues are given by
\begin{equation}
E_n = -{1\over {2}}{\left[ \nu/2-1-n\right]}^2, \quad\quad\quad n=0,1,\dots 
[\nu/2-1]
\label{bev}
\end{equation}
where $[m]$ indicates the integer part of $m$. Clearly this discrete spectrum exists provided $\nu>2$.    
The corresponding normalized eigenfunctions are given by
\begin{equation}
g_n(\phi)= b_n e^{-\phi/2} W_{ {{\nu-1}\over {2}}, {{\nu}\over {2}}-1-n}\left(2ae^{\phi}\right),
\label{bef}
\end{equation}
where $W_{\alpha,\beta}(x)$ is the Whittaker function\cite{GR}. The normalization constant $b_n$
can also be computed exactly (see appendix A)
\begin{equation}
b_n^2=\frac{2\sigma (\nu-2-2n)\Gamma(n+1)}{\Gamma(\nu-1-n)}
\label{bnorm}
\end{equation}
where 
$\Gamma(x)$ is the standard Gamma function.  
\vspace{\baselineskip}

\noindent {\bf Scattering States:} The scattering states have positive energies labelled by the wavevector
$q$, $E_q = q^2/2$ with $0\le q\le \infty$. The corresponding eigenfunctions are given by
\begin{equation}
g_q(\phi)=b(q) e^{-\phi/2} W_{ {{\nu-1}\over {2}}, iq}\left(2a e^{\phi}\right),
\label{sef}
\end{equation}
where the coefficient $b(q)$ is given by
\begin{equation}
b(q)= {1\over {\sqrt{2\pi}}} (2a)^{-iq-1/2} { {\Gamma(1-\nu/2-iq)}\over {\Gamma(-2iq)}}.  
\label{bq}
\end{equation}
This coefficient $b(q)$ is chosen such that in the limit $\phi\to -\infty$ (where the quantum potential
in the Hamiltonian $\hat H_M$ in Eq. (\ref{morse}) vanishes) the eigenfunction $g_q(\phi)$ approaches
a plane wave form, i.e., $g_q(\phi)\to {1\over {\sqrt{2\pi}}}\left[e^{iq\phi}+r(q)e^{-iq\phi}\right]$
as $\phi\to -\infty$, where $e^{iq\phi}$ represents an incident wave travelling in the direction of 
positive $\phi$ and $e^{-iq\phi}$ represents the reflected wave travelling in the opposite direction 
with $r(q)$ being the reflection coefficient (for details see appendix-A).

Having obtained the full spectrum of ${\hat H_M}$ we are now ready to compute the persistence profile 
${\overline {P(x)}}=1-{\overline {Q_0(x)}}$. Expanding the right hand side of 
Eq. (\ref{compact1}) in the energy basis of ${\hat H_M}$ and using the results on the spectrum of 
${\hat H_M}$ summarized above, we get
\begin{equation}
{\overline {Q_0(x)}}= e^{-\sigma \nu^2 x/2}\left[ \sum_{n=0}^{[\nu/2-1]}|<g_n|\psi_0>|^2 e^{2\sigma(\nu/2-1-n)^2 x}
+\int_0^{\infty} dq |<g_q|\psi_0>|^2 e^{-2q^2\sigma x}\right],
\label{profileq1}
\end{equation}
a result which is valid for all $x$ and for all $\mu>0$. The function $\psi_0(\phi)$ is already known. In fact,
for $p=0$ we find from Eqs. (\ref{bphi}) and (\ref{psi00}) the following normalized expression
\begin{equation}
\psi_0(\phi)= {1\over {\sqrt{ \Gamma(\nu)(2\sigma)^\nu}}}\exp\left[-{1\over {4\sigma}}e^{\phi}+ {\nu\over 
{2}}\phi\right].
\label{psi0}
\end{equation}
Using this expression of $\psi_0(\phi)$ and the eigenfunctions in Eqs. (\ref{bef}) and (\ref{sef}) one can
easily evaluate the matrix elements $<g_n|\psi_0>$ and $<g_q|\psi_0>$. For the bound states we get
\begin{equation}
<g_n|\psi_0>= \int_{-\infty}^{\infty} d\phi \psi_0(\phi)g_n(\phi)= { {b_n}\over {\sqrt{2\sigma \Gamma(\nu)}} 
}\Gamma(n+1)\Gamma(\nu-1-n),
\label{bmatrix}
\end{equation}
where $b_n$ is given by Eq. (\ref{bnorm}). Similarly for the scattering states we obtain
\begin{equation}
<g_q|\psi_0>= \int_{-\infty}^{\infty} d\phi \psi_0(\phi)g_q(\phi)=  {{b(q)}\over {\sqrt{2\sigma 
\Gamma(\nu)}}}\Gamma\left(\nu/2-iq\right)\Gamma\left(\nu/2+iq\right),
\label{smatrix}
\end{equation}
with $b(q)$ given by Eq. (\ref{bq}). Substituting these matrix elements in Eq. (\ref{profileq1}) 
we get our final expression,
\begin{equation}
{\overline {Q_0(x)}}= {{e^{-\sigma \nu^2 x/2}}\over {2\sigma \Gamma(\nu)}}\left[ \sum_{n=0}^{[\nu/2-1]} b_n^2 
\Gamma^2(n+1)\Gamma^2(\nu-1-n) 
e^{2\sigma(\nu/2-1-n)^2 x}
+\int_0^{\infty} dq {|b(q)|}^2 {|\Gamma\left(\nu/2-iq\right)|}^4 
e^{-2q^2\sigma x} \right],
\label{profileq2}
\end{equation} 
where $b_n$ and $b_q$ are given respectively by Eqs. (\ref{bnorm}) and (\ref{bq}). 
Substituting Eq. (\ref{profileq2}) in the relation ${\overline {P(x)}}=1-{\overline {Q_0(x)}}$
then gives us the exact persistence profile valid for any $x>0$ and any $\mu>0$.

It is instructive to derive explicitly the tails of this profile ${\overline {P(x)}}$ for small $x$
and large $x$. Consider first the limit $x\to 0$. While we can use the general solution in Eq. (\ref{profileq2})
to derive the small $x$ behavior, it is easier to consider the original equation (\ref{compact}) which
for small $x$ gives ${\overline {Q_0(x)}}\to 1- 4\sigma x <\psi_0|{\hat H}|\psi_0>$. 
Since ${\hat H}={\hat H_0}+{\hat H_1}$
and moreover since $\psi_0(\phi)$ is a zero energy eigenfunction of ${\hat H_0}$, we get
${\overline {Q_0(x)}}\to 1- 4\sigma x <\psi_0|{\hat H_1}|\psi_0>$. Expanding the matrix element in the 
$\phi$-basis,
using ${\hat H_1}=a e^\phi$ (setting $\lambda=1$ in Eq. (\ref{hamil})) and the expression of $\psi_0(\phi)$
from Eq. (\ref{psi0}) and evaluating the resulting integral, we get ${\overline {Q_0(x)}}\to 1- 2\sigma \nu 
x$.
Using $\nu=\mu/\sigma$, we get ${\overline {P(x)}}=2\mu x$ as $x\to 0$. Thus we obtain an
interesting result that the slope $2\mu$ characterizing the linear growth of the profile near
$x=0$ is completely independent of the disorder strength $\sigma$. 
In fact, the small $x$ behavior of the disorder averaged persistence profile is identical
to that of the pure case. 

We now turn to the other limit $x\to \infty$. Here we use Eq. (\ref{profileq2}). First consider
the case when $\nu>2$. Then we know from Eq. (\ref{profileq2}) that there exist bound states.
In that case it is evident that for large $x$, the term corresponding to the lowest energy bound state ($n=0$)
will be the most dominant term on the right hand side of Eq. (\ref{profileq2}). Retaining only this leading 
$n=0$
term in Eq. (\ref{profileq2}) and using $b_0^2=2\sigma/\Gamma(\nu-2)$ we get, ${\overline 
{Q_0(x)}}={{\nu-2}\over {\nu-1}}e^{-2\sigma (\nu-1)x}$
as $x\to \infty$ for $\nu>2$. Consider now the opposite case when $\nu<2$. In this case there are no bound states
and there is no contribution from the discrete sum on the right hand side of Eq. (\ref{profileq2}).
The only contribution is from the integral representing the scattering states. For large $x$, the
most dominant constribution to the integral will come from the small $q$ regime. Expanding
the Gamma functions for small $q$, we find after preliminary algebra, ${\overline {Q_0(x)}}=
A_\nu (2\sigma x)^{-3/2}e^{-\nu^2\sigma x/2}$ as $x\to \infty$ for $\nu<2$ where $A_\nu$ is a constant (see below).
Exactly at $\nu=2$, we get from Eq. (\ref{profileq2}), ${\overline {Q_0(x)}}=e^{-2\sigma x}/\sqrt{2\pi 
\sigma 
x}$ for large 
$x$. Let us summarize the three different types of large $x$ behaviors of the persistence profile,
\begin{equation}
1-{\overline {P(x)}} \to\cases{
                 {{\nu-2} \over {\nu-1}}e^{-2(\nu-1)\sigma x}   &$\nu>2$,\cr
                 {1\over {\sqrt{2\pi \sigma x}}}e^{-2\sigma x}           &$\nu=2$, \cr
                 {A_{\nu}\over {(2\sigma x)^{3/2}} }e^{-\nu^2 \sigma x/2} &$\nu<2$,\cr}
\label{profilef}
\end{equation}
where $A_{\nu} = {\pi^{3/2}\Gamma^2(\nu/2)}/{\Gamma(\nu)\left[1-\cos (\pi\nu)\right]}$.
Evidently the shape of the profile in Eq. (\ref{profilef}) for large $x$
changes as $\nu$ varies through the critical point $\nu_c=2$.
The reason for the existence of this critical point is evident from our analysis.
Essentially it happens due to the loss of bound states as $\nu$ decreases from $\nu>2$
to $\nu<2$. Note that this critical behavior at finite $\nu=\nu_c=2$ could not be 
derived by the RSRG method.
The RSRG method is valid only in the $\nu\to 0$ limit where the exact result in Eq. (\ref{profilef})
coincides with the RSRG results\cite{LMF}.                                          

We conclude this section by pointing out that it is possible to have an alternative derivation 
of the disorder averaged persistence profile ${\overline {P(x)}}$ for $\mu>0$ by a completely
different method. This method relies on mapping the calculation of the persistence 
profile to calculating the disorder average of the ratio of two partition functions
in the Sinai model. This mapping  
makes use of certain mathematical properties of the Brownian motion.
The average of this ratio of partition functions was already computed before\cite{KT1,CMY}
in a different context. Using those results one can then recover the results in Eq. (\ref{profilef}).
The derivation of this mapping is presented in appendix-B.

\section{Explicit Results for negative Drift $(\mu<0)$}   

We now turn to the the case when $\mu<0$ which corresponds to a drift towards
the origin since the initial position $x>0$. The situation here is very different
from the positive drift $\mu>0$ case discussed in the previous section. For $\mu<0$,
we expect from the analogy to the pure case that for each sample, the particle will
definitely cross the origin as $t\to \infty$ no matter what the starting position $x$ is.
Hence for $\mu<0$, the persistence $P(x,t)\to 0$ as $t\to \infty$ for all $x$, unlike
the $\mu>0$ case where the persistence approaches a time independent profile $P(x,t)\to 
P(x)$ as $t\to \infty$. Therefore what is interesting in the $\mu<0$ case is to 
compute the asymptotic behavior of $P(x,t)$ for large $t$. Recalling
the definitions $P(x,t)=1-\int_0^{t}F(x,t_*)dt_*$ and
$Q_p(x)=\int_0^{\infty}e^{-pt}F(x,t)dt$ where $F(x,t)$ is the first-passage time
distribution, we see that the analysis of the large but finite $t$ limit
of the disorder averaged persistence ${\overline {P(x,t)}}$ requires an analysis of 
the $p\to 0$ limit of the Laplace transform ${\overline {Q_p(x)}}$ [rather
than exactly at $p=0$ where ${\overline {Q_0(x)}}=1$ trivially for $\mu<0$]. 
In fact we will see later in this section that in this limit, ${\overline {P(x,t)}}$
or equivalently ${\overline {F(x,t)}}$ display a variety of scaling behaviors 
as one tunes the relevant parameter $\nu'= -\mu/\sigma$.

As in the case of $\mu>0$, the starting point of our analysis for $\mu<0$ is the central 
result in Eq. (\ref{compact}) which is valid for all $\mu$. Unlike the $\mu>0$ case, we
can not, however, put $p=0$ straightway in the Hamiltonian ${\hat H}$ in Eq. (\ref{hamil}).
To derive the time dependent asymptotics for large $t$, we need to analyze the
spectrum of ${\hat H}$ keeping $p$ small but {\em nonzero}. Unfortunately, for nonzero
$p$, it is hard to determine the full spectrum of ${\hat H}$ exactly. Fortunately,
however, it is possible to extract the leading asymptotic behavior as $x\to \infty$
and $p\to 0$ without too much trouble. To see this, we consider the energy eigenvalue
decomposition in the second line on the right hand side of Eq. (\ref{compact})
and find that as $x\to \infty$, the leading contribution comes from the ground state
energy $E_0$ of ${\hat H}$,
\begin{equation}
{\overline {Q_p(x)}}\to {|<\psi_0|g_0>|}^2 e^{-4\sigma E_0 x},
\label{gs1}
\end{equation}
where $|g_0>$ is the ground state of ${\hat H}$. 
It is clear
from Eq. (\ref{gs1}) that to evaluate the large $x$ asymptotics we just need to
compute the ground state energy $E_0$ of the Hamiltonian ${\hat H}$
in Eq. (\ref{hamil}).
Unfortunately, even the ground state energy $E_0$ is hard to compute exactly for arbitrary $p$.
One can, however, make progress in the $p\to 0$ limit. To see this we  
recall that $|\psi_0>$ in Eq. (\ref{gs1}) is the 
exact zero energy eigenstate of the `free' part of the Hamiltonian ${\hat H_0}$ (see the
discussion after Eq. (\ref{coeff})), i.e., ${\hat H_0}|\psi_0>=0$. Knowing this exact
fact, we can then determine the ground state energy $E_0$ of the full Hamiltonian
 ${\hat H}={\hat H_0}+{\hat H_1}$ by treating ${\hat H_1}=\lambda a e^{\phi}$ (where
eventually we will set $\lambda=1$) as a perturbation to the `free' Hamiltonain ${\hat H_0}$. 
For example, to first order
in the perturbation term, we get $E_0 = 0 + <\psi_0|{\hat H_1}|\psi_0>$ where $0$
indicates the fact the ground state energy of the unperturbed Hamiltionain ${\hat H_0}$
is exactly $0$. We then decompose this matrix element $<\psi_0|{\hat H_1}|\psi_0>$ in the
$\phi$ basis and use ${\hat H_1}=ae^{\phi}$ (setting $\lambda=1$) to obtain
$E_0=a\int_{-\infty}^{\infty} \psi_0^2(\phi) e^{\phi}d\phi$. The normalized 
wavefunction $\psi_0 (\phi)$ is already known from Eq. (\ref{psi00}) and
has the following explicit expression,
\begin{equation}
\psi_0(\phi)= \sqrt{ { {(2p)^{\nu'/2}}\over {2K_{\nu'}\left(\sqrt{2p}/\sigma\right)}}}
\exp\left[-{e^{\phi}\over {4\sigma}}-{\nu'\over {2}}\phi - {p\over {2\sigma}}e^{-\phi}\right],
\label{psi01}
\end{equation}
where $K_{\alpha}(x)$ is the modified Bessel function of index $\alpha$\cite{GR} and we have used
the definition $\nu'=-\mu/\sigma$. Using this explicit form of $\psi_0(\phi)$ and carrying out
the integral in $E_0=a\int_{-\infty}^{\infty} \psi_0^2(\phi) e^{\phi}d\phi$ we get,
to first order in perturbation theory,
\begin{equation}   
E_0 =a\sqrt{2p} { {K_{1-\nu'}\left(\sqrt{2p}/\sigma\right)}\over
{K_{\nu'}\left(\sqrt{2p}/\sigma\right)} }.
\label{e0}
\end{equation}

The result in Eq. (\ref{e0}) is only upto first order in ${\hat H_1}$. It is not clear, so far, why
one should stop at first order only. In other words, we have not yet specified in what sense
${\hat H_1}$ is `small' compared to ${\hat H_0}$. Note that in deriving Eq. (\ref{e0}) we have
not yet taken the $p\to 0$ limit. In the limit $p\to 0$, using the asymptotic properties of Bessel 
functions\cite{GR} in Eq. (\ref{e0}), we find
\begin{equation}
E_0 \to\cases{
                 {a \over {\sigma(\nu'-1)}}p         &$\nu'>1$,\cr
                 -{a\over {\sigma }}p\log p           &$\nu'=1$, \cr
                 a B_{\nu'}p^{\nu'}                     &$0<\nu'<1$,\cr}
\label{e01}
\end{equation}    
where we recall that $a=1/{4\sigma}$ and the constant $B_{\nu'}$ is given by
\begin{equation}
B_{\nu'}={{2^{1-\nu'}\Gamma(1-\nu')}\over {\sigma^{2\nu'-1}\Gamma(\nu')}}.
\label{Bnu}
\end{equation}
Thus, in general, $E_0\sim p^{\alpha}$ as $p\to 0$ where $\alpha=1$ for $\nu'\ge 1$ and $\alpha=\nu'$ 
for $\nu'\le 1$ with additional logarithmic corrections at $\nu'=1$. Hence basically $E_0$ is `small' for small
$p$. Naively one would expect that if, indeed, we carry out the perturbation theory in ${\hat H_1}$
to higher orders, the resulting terms will be lower order in $p$ as $p\to 0$. This naive expectation, fortunately,
turns out to be true. In fact we show in appendix C how to estimate the second order term for small $p$
and it turns out to be at least of $O(p^{\alpha+1})$ and hence negligible compared to the first order term 
($\sim O(p^{\alpha})$)
in the $p\to 0$ limit. For small $p$, this argument therefore justifies in keeping only the first order term in the
perturbation theory in evaluating $E_0$. 
Note that the eigenfunction $|g_0>$ also gets modified
from the `free' eigengunction $|\psi_0>$ due to the perturbation term.  

Substituting the $p\to 0$ results from Eq. (\ref{e01}) in Eq. (\ref{gs1}) for large $x$ and using
$a=1/{4\sigma}$, we then get
three different types of scaling behaviors depending on $\nu'$,
\begin{equation}
{\overline {Q_p(x)}} \to\cases{
                  \exp\left[-px/{\sigma(\nu'-1)}\right]  &$\nu'>1$,\cr
                  \exp\left[\, p\log (p)x/\sigma\right]     &$\nu'=1$, \cr
                  \exp\left[-B_{\nu'}p^{\nu'}x\right]    &$\nu'<1$,\cr}
\label{ndr1}
\end{equation}
where $B_{\nu'}$ is given by Eq. (\ref{Bnu}). Note that to
leading order in small $p$,
we need to keep only the zeroth order term in the amplitude $|<\psi_0|g_0>|^2$ of the exponential in 
Eq. (\ref{gs1}). To the zeroth order $|g_0>=|\psi_0>$ and hence to leading order this amplitude is
exactly $1$ since $|\psi_0>$ is normalized. Thus it is evident from Eq. (\ref{ndr1}) that for $\nu'>1$, the 
correct scaling limit
is $x\to \infty$, $p\to 0$ but keeping the product $px$ fixed. On the other hand, for $\nu'<1$,
the correct scaling combination is $p^{\nu'}x$.  

It turns out that the exact asymptotic results in Eq. (\ref{ndr1}) were also derived recently by 
mathematicians by using completely different methods which involved rather heavy mathematical machineries. 
The first
derivation is due to Kawazu and Tanaka who used the so called Kotani's formula and Krein's
theory of strings\cite{KT2,Tan}. However their method didn't permit the explicit evaluation of the
constant $B_{\nu'}$. More recently, Hu et. al. \cite{HSY} presented yet another completely
different derivation by mapping the persistence problem with negative drift onto that of a
Bessel process and then using some theorems on this Bessel process. Hu et. al. managed to compute 
the coefficient $B_{\nu'}$ explicitly. Note, however, that the constant $B_{\nu'}$ in ref. \cite{HSY}
has an apparently rather different looking form than our expression in Eq. (\ref{Bnu}) and it requires a bit or work
to show that indeed they are exactly identical. 

To derive the asymptotic properties of the first-passage time distribution $F(x,t)$ in the time domain, we need to 
invert the Laplace transform in Eq. (\ref{ndr1}) with respect to $p$. For $\nu'>1$, the Laplace inversion is trivial and
we get
$F(x,t)=\delta \left(t- x/{\sigma (\nu'-1)}\right)$ in the scaling limit $x\to \infty$, $t\to \infty$ but keeping
the ratio $x/t$ fixed. This result is very similar to the
pure case. The delta function indicates that at late times the particle
basically moves ballistically with an effective velocity $\sigma(\nu'-1)$. Similarly for $\nu'=1$, by inverting
the Laplace tranform we find that in the scaling limit $x\to \infty$, $t\to \infty$ but keeping the ratio
$x\log (x)/t$ fixed, $F(x,t)=\delta (t-x\log(x)/\sigma)$. The situation becomes somewhat different
for $\nu'<1$. In this case the Laplace inversion indicates that in the scaling limit $x\to \infty$, $t\to \infty$
but keeping the
ratio $x^{1/\nu'}/t$ fixed, the first-passage time distribution approaches a scaling form
$F(x,t)\sim {1\over t}f(t/x^{1/\nu'})$. The scaling function $f(y)$ can be formally written in terms of 
L\'evy function $L_{\nu'}(y)$ and we get $f(y)=yL_{\nu'}(y)$. The L\'evy function is 
formally defined by the Bromwich integral (see the 
appendix 
of ref. \cite{BG}),
\begin{equation}
L_{\nu'}(y)= {1\over {2\pi i}}\int_{d-i\infty}^{d+i\infty} ds e^{sy-B_{\nu'}s^{\nu'}},
\label{levy}  
\end{equation}
where $d$ is such that the $s=d$ line in the complex $s$ plane lies to the right of all the singularities
(branch cuts) but is otherwise arbitrary. 

An explicit expression of the L\'evy function can be obtained only for a few special values of 
$\nu'$\cite{BG}. 
For example, for $\nu'=1/2$, we get $f(y)=yL_{1/2}(y)= e^{-1/2y}/\sqrt{2\pi y}$. This indicates 
that $F(x,t)= xe^{-x^2/4t}/\sqrt{2\pi t^3}$ and hence the persistence 
$P(x,t)= 1-\int_0^{t} F(x,t')dt'={\rm erf}(x/\sqrt{2t})$.
We thus arrive at an amazing result that for $\nu'=1/2$, i.e., $\mu=-\sigma/2$, the {\em disorder averaged}
persistence has the same asymptotic behavior as the {\em pure case without drift} ($\mu=0$, $\sigma=0$) (as derived in
section II-A)! A similar coincidence at this special value of $\nu'=1/2$ was also noted in the context of
occupation time distribution in the Sinai model\cite{MC}. Thus $\mu=-\sigma/2$ seems to be a 
special line in the $(\mu-\sigma)$ plane where the Sinai model with drift shares the same
asymptotic properties as the pure unbiased Brownian motion. Another solvable point is $\nu'=1/3$,
where we get $ f(y)=y L_{1/3}(y)= {{\alpha_0}\over {2\pi \sqrt{y}}}K_{1/3}\left(\alpha_0/\sqrt{y}\right)$
with the constant $\alpha_0= {4\over {3^{3/2}}}{\left[ {{\Gamma(2/3)}\over {\Gamma(1/3)}}\right]}^{3/2}\sqrt{\sigma}$.

For general $\nu'<1$, while we can not calculate the scaling function $f(y)$ explicitly, the behaviors
at the tails can be easily determined. For example, first consider the limit $y=t/x^{1/\nu'}\to \infty$, i.e., when
$t>> x^{1/\nu'}$. Using the large $y$ behavior of the L\'evy function\cite{BG},
$L_{\nu'}(y)\approx B_{\nu'}\Gamma(1+\nu'){\sin (\pi \nu')}/{\pi y^{\nu'+1}}$, we get
$F(x,t)\approx \beta_0 x/{t^{\nu'+1}}$ where $\beta_0={\nu' 2^{1-\nu'}}/{\Gamma(\nu')\sigma^{2\nu'-1}}$.
The persistence $P(x,t)=1-\int_0^t F(x,t')dt'$
then behaves in this limit as
\begin{equation}
P(x,t) \approx { {2^{1-\nu'} }\over {\Gamma(\nu')\sigma^{2\nu'-1} } } {x\over {t^{\nu'}}}, \quad\quad\quad t>>x^{1/\nu'},
\label{pers1}
\end{equation}
indicating a power law decay $P(x,t)\sim t^{-\theta}$ for large $t$ where the persistence exponent $\theta=\nu'$.
We now turn to the other tail of the scaling function $f(y)$ when $y=t/x^{1/\nu'}\to 0$, i.e., when $t<< x^{1/\nu'}$.
Using the properties of the L\'evy
function near $y\to 0$\cite{BG}, we get
$f(y)=yL_{\nu'}(y)\approx e^{-(1-\nu')/{\nu'\zeta}}/\sqrt{2\pi (1-\nu')\zeta}$ where $\zeta=[y^{\nu'}/{\nu'
B_{\nu'}}]^{1/(1-\nu')}$. This indicates an essential singularity at $y=0$. Using this 
asymptotic behavior of $F(x,t)$ in the relation $P(x,t)=1-\int_0^t F(x,t')dt'$ we find the
following behavior for the persistence,
\begin{equation}
P(x,t)\approx 1- {1\over {\sqrt{2\pi \beta_1(1-\nu')}}}\left({x\over
{t^{\nu'}}}\right)^{-1/{2(1-\nu')}}\exp\left[-{{1-\nu'}\over {\nu'}}\beta_1 \left({x\over
{t^{\nu'}}}\right)^{1/(1-\nu')}\right], \quad\quad\quad t<< x^{1/\nu'},
\label{pers2}
\end{equation}
where $\beta_1= (\nu'B_{\nu'})^{1/(1-\nu')}$ and $B_{\nu'}$ is given by
Eq. (\ref{Bnu}).

Let us summarize the main behavior of the disorder averaged persistence for $\nu'<1$.  We find that for 
$\nu'<1$, there 
exists a single time scale $t_x\sim x^{1/\nu'}$ depending on the initial position $x$.
For $t>>t_x$, the persistence $P(x,t)$ decays as a power law with an exponent $\theta=\nu'$
and the amplitude of the power law depends on $x$ as in Eq. (\ref{pers1}). In the opposite limit
when $t<< t_x$, the persistence drops extremely sharply from its initial value $P(x,0)=1$
as indicated by the essential singularity in the second term on the right hand side
of Eq. (\ref{pers2}) when $t<< x^{1/\nu'}$.

Let us conclude this section on negative drift by one final remark. We note that even though we
have assumed throughout this section that $\mu$ is strictly negative, we can safely take
the limit $\mu\to 0^{-}$ in Eq. (\ref{e0}) which gives $E_0\approx -1/{[2\log p]}$ in
the limit $p\to 0$. It then follows from Eq. (\ref{gs1}) that
in the limit of vanishing drift, one gets the asymtotic result ${\overline {Q_p(x)}}\to
1 +
{{2\sigma}\over {\log p}}x +\ldots$ when $x<< -1/{\log p}$. From this it follows that
${\overline {P(x,t)}}\approx {2\sigma x}/\log t$ for $\log t>> x$, thus
recovering the standard Sinai model behavior in the zero drift limit\cite{LMF,CD}.

\section{Summary and Perspectives}

In summary, we have obtained exact asymptotic results for the disorder averaged
persistence in the Sinai model with an arbitrary drift. Our method maps exactly
the problem of computing the persistence to the problem of finding
the eigenvalue spectrum of a single particle quantum Hamiltonian. We have shown
that it is possible to find this spectrum in certain cases which allowed
us to obtain exact asymptotic results for arbitrary drift. We note that
these results could not have been obtained from the existing
physical methods (e.g. the RSRG method) which provide exact 
results only in the limit of zero drift.
Our results show that there is a rich variety of asymptotic behaviors in
the persistence as one tunes the drift. In particular, the asymptotics 
undergo interesting `phase transitions' at certain critical values of the control
parameter $\nu$ (the relative strength of the drift over the disorder), e.g.,
at $\nu=2$, $\nu=0$ and $\nu=-1$. 
It would be interesting to extend the exact method presented here to calculate
other properties in the Sinai model with finite drift, such as the persistence of a thermally 
averaged trajectory for which the results in the zero drift limit are already known\cite{LMF,IR}.

We thank D.S. Dean for useful discussions.

\appendix

\section{Derivation of the spectrum of the Hamiltonian $\hat H_M$}

In this appendix we derive the spectrum of the generalized Morse Hamiltonian ${\hat H_M}$ given
by Eq. (\ref{morse}). We show that the spectrum has a discrete part with negative energies
which correspond to the bound states and also a continuous part with positive energies
corresponding to scattering states. The eigenfunctions satisfy the
Schr\"odinger equation,
\begin{equation}
-{1\over {2}} {{d^2 g_E(\phi)}\over {d \phi^2}}+ \left[{a^2\over {2}}e^{2\phi}-
{{a(\nu-1)}\over {2}}e^{\phi}\right]g_E(\phi)=E g_E(\phi).
\label{ashrodinger}
\end{equation} 
Let us remind the readers that $a=1/{4\sigma}$. It turns out to be convenient to make a change of variable 
$y=2a e^{\phi}$. Furthermore, let
us substitute $g_E(\phi)= e^{-\phi/2}f(2ae^{\phi})$ in Eq. (\ref{ashrodinger}). 
Then the function $f(y)$ satisfies the differential equation,
\begin{equation}
{ {d^2f}\over {dy^2}} +\left[-{1\over {4}}+ { {(\nu-1)}\over {2y}} + { {1/4+2E}\over {y^2}}\right]f(y)=0,
\label{fdeq}
\end{equation}
where we have suppressed the $E$ dependence of the function $f(y)$ for notational
convenience. We next consider the negative and the positive part of the spectrum separately.
\vspace{\baselineskip}

\noindent {\bf Bound States:} The bound states are in the negative energy part of the 
spectrum. Let us substitute 
$E=-\gamma^2/2$ in Eq. (\ref{fdeq}) where $\gamma$ is the eigenvalue to be determined.
The second order differential equation (\ref{fdeq}) with $E=-\gamma^2/2$ is known to
have two linearly independent solutions $ W_{ {{\nu-1}\over {2}}, \gamma}(y)$
and $W_{ -{{\nu-1}\over {2}}, \gamma}(-y)$ where $W_{\alpha,\beta}(x)$ is the Whittaker function\cite{GR}.     
Thus the most general solution of Eq. (\ref{fdeq}) can be written as
\begin{equation}
f(y)=D_1 W_{ {{\nu-1}\over {2}}, \gamma}(y) + D_2 W_{ -{{\nu-1}\over {2}}, \gamma}(-y),  
\label{fsol}
\end{equation}
where $D_1$ and $D_2$ are 
arbitrary constants. For large argument, the Whittaker
function is known to have the asymptotic behavior\cite{GR}, $W_{\alpha,\beta}(x)\sim x^{\alpha}e^{-x/2}$.
On the other hand the bound states must be normalizable and hence the eigenfunction
$g_E(\phi)$ must vanish as $\phi\to \pm \infty$. The vanishing boundary condition at 
$\phi=\infty$ indicates that the constant $D_2=0$. Note that here we have assumed $\gamma>0$.
If $\gamma<0$, then this boundary condition instead sets $D_1=0$. However, the resulting
solution is the same. In other words the eigenfunctions corresponding to $\gamma $ and $-\gamma$
are the same and not linearly independent of each other. Thus without any loss of generality
we can assume $\gamma>0$ and set $D_2=0$. Thus the 
eigenfunction in terms of
the original variable is given by
\begin{equation}
g_E(\phi)= D_1 e^{-\phi/2}  W_{ {{\nu-1}\over {2}}, \gamma}(2a e^\phi).
\label{beigen1}
\end{equation}
Note that the eigenvalue $\gamma$ is
yet to be determined. This is done by employing the vanishing boundary condition at the other tail,
namely $g_E(\phi)\to 0$ as $\phi\to -\infty$. Using the small $x$ behavior of the Whittaker function,
$W_{\alpha,\beta}(x)\to { {\Gamma(2\beta)}\over {\Gamma(1/2+\beta-\alpha)}}x^{-\beta+1/2}$ as $x\to 0$
in Eq. (\ref{beigen1}) as $\phi\to -\infty$ we get
\begin{equation}
g_E(\phi)\approx D_1 (2a)^{-\gamma+1/2}{ {\Gamma(2\gamma)}\over 
{\Gamma(1-\nu/2+\gamma)}}e^{-\gamma \phi}.
\label{beigen2}
\end{equation} 
Thus the eigenfunction diverges exponentially as $\phi\to -\infty$. The only way the eigenfunction
can satisfy the boundary condition $g_E(-\infty)=0$ is if the denominator $\Gamma(1-\nu/2-\gamma)$
in Eq. (\ref{beigen2}) is infinite. This happens when the argument of the Gamma function is a
negative integer $1-\nu/2-\gamma=-n$ with $n=0,1,\dots$. This thus fixes the eigenvalue $\gamma=\nu/2-1-n$
with $n=0,1,\dots$. However note that the condition $\gamma>0$ indicates that the maximal 
allowed value for $n$ is $[\nu/2-1]$ where $[x]$ denotes the integer part of $x$. Thus finally 
the bound states have discrete eigenvalues $E= -\gamma^2/2=-(\nu/2-1-n)^2/2$ and the corresponding
eigenfunctions, labelled by $n$, are given by
\begin{equation}
g_n(\phi)= b_n e^{-\phi/2} W_{ {{\nu-1}\over {2}}, {{\nu}\over {2}}-1-n}\left(2ae^{\phi}\right),
\label{beigen3}
\end{equation}   
where $b_n=D_1$ is to be fixed from the normalization condition, $\int_{-\infty}^{\infty} 
g_n^2(\phi)d\phi=1$. To perform this integral,
we first use the fact that for positive integer $n$,
one can rewrite the Whittaker function in terms of the Laguerre polynomials 
\cite{GR} and then use the following identity (see the  appendix of \cite{CH})
\begin{equation}
\int_0^{\infty} dx x^{\alpha-1}\, e^{-x}\left[L_{n}^{\alpha}(x)\right]^2 =
\frac{\Gamma(\alpha+n+1)}{\alpha\Gamma(n+1)}.
\label{laguerre}
\end {equation}
One obtains
\begin{equation}
b_n^2=\frac{2\sigma (\nu-2-2n)\Gamma(n+1)}{\Gamma(\nu-1-n)}.
\label{intid}
\end{equation}

\vspace{\baselineskip}

\noindent{\bf Scattering States:} We now turn to the positive energy part of the specrum
and set $E=q^2/2$ in Eq. (\ref{fdeq}). The resulting differential equation, once again, has 
two linearly independent solutions $ W_{ {{\nu-1}\over {2}}, iq}(y)$
and $W_{ -{{\nu-1}\over {2}}, iq}(-y)$. Note that the function 
$W_{ -{{\nu-1}\over {2}}, iq }(-y)= W_{ -{{\nu-1}\over {2}}, iq}(-2ae^\phi)\sim \exp(e^\phi)$ as
$\phi\to \infty$. Since the eigenfunction $g_E(\phi)$, even though non-normalizable for scattering states,
can not diverge superexponentially as $\phi\to \infty$, this second solution is not allowed.
Keeping only the first solution we get the eigenfuntions, now labelled by $q$,
\begin{equation}
g_q(\phi)= b(q) e^{-\phi/2} W_{ {{\nu-1}\over {2}}, iq}\left(2a e^{\phi}\right).
\label{seigen1}
\end{equation}
The question is how to determine the constant $b(q)$ in Eq. (\ref{seigen1}). This is because, unlike the 
bound states, the eigenfunctions in Eq. (\ref{seigen1}) are non-normalizable. To see this let us
examine the behavior of $g_q(\phi)$ near the tail $\phi\to -\infty$, as in the discrete case.
Using the asymptotic properties of the Whittaker function we find that as 
$\phi\to -\infty$,
\begin{equation}
g_q(\phi)\to b(q)\left[ { {\Gamma(-2iq)}\over {\Gamma(1-\nu/2-iq)}}(2a)^{iq+1/2}e^{iq\phi}
+ {{\Gamma(2iq)}\over {\Gamma(1-\nu/2+iq)}}(2a)^{-iq+1/2}e^{-iq\phi}\right].
\label{seigen2}
\end{equation}
Clearly the functions $g_q(\phi)$'s are non-normalizable. Moreover, unlike in the case
of bound states where the boundary condition at $\phi=-\infty$ decides the discrete
eigenvalues, in this case we have no such condition indicating that all possible values of $q\ge0$
are allowed. Note that, as in the discrete case, $q>0$ and $q<0$ correspond to the
same eigenfunction and hence the allowed values of $q$ lie in the range $0\le q\le \infty$.
To determine the constant $b(q)$ we note that in the tail $\phi\to -\infty$, the quantum potential
in Eq. (\ref{ashrodinger}) vanishes. The resulting differential equation with $E=q^2/2$
allows plane wave solutions of the form
\begin{equation}
g_q(\phi)\approx {1\over {\sqrt{2\pi}}}\left[e^{iq\phi}+r(q)e^{-iq\phi}\right],
\label{planew}
\end{equation}
where $e^{iq\phi}$ represents the incoming wave coming from $\phi= -\infty$
and $e^{-iq\phi}$ represents the reflected wave going towards $\phi=-\infty$
with $r(q)$ being the reflection coefficient. The amplitude $1/\sqrt{2\pi}$
ensures that the plane waves $\psi_q(x)={1\over {\sqrt{2\pi}}}e^{iqx}$ 
are properly ortho-normalized in the sense that $<\psi_q|\psi_q'>=\delta(q-q')$
where $\delta(z)$ is the Dirac delta function. Comparing Eqs. (\ref{seigen2})
and (\ref{planew}) in the regime $\phi\to -\infty$, we determine the constant
$b(q)$ up to a phase as
\begin{equation}
b(q)= {1\over {\sqrt{2\pi}}} (2a)^{-iq-1/2} { {\Gamma(1-\nu/2-iq)}\over {\Gamma(-2iq)}}.
\label{abq}
\end{equation}    
  
This completes the derivation of the spectrum of the Hamiltonian ${\hat H_M}$.

\section{Alternative Derivation of the Persistence Profile for $\mu>0$}

For each sample of the disorder, the persistence profile $P(x)$ is related 
to the Laplace transform $Q_p(x)$ of the first-passage time distribution
at $p=0$ via the relation, $P(x)=1-Q_0(x)$. 
The quantity $Q_0(x)$ can be obtained exactly by solving the
Eq. (\ref{bfp}) with $p=0$,
\begin{equation}
Q_0(x)=1- {{Z_{\mu}(x)}\over {Z_{\mu}(\infty)}},
\label{profile}
\end{equation}
where $Z_{\mu}(x)= \int_0^{x} e^{2U(x')}dx'$ is the partition function in a finite box of size $x$
with $U(x)=-\int_0^x F(x')dx'= -\mu x +\sqrt{\sigma} B(x) $  being the random potential. It turns out to be 
useful to rewrite Eq. (\ref{profile}) in a slightly different form using a well known 
property of the Brownian motion: If $B(x)$ is a Brownian motion, then 
$B(x)-B(x')\equiv {\tilde B}(x-x')$ where ${\tilde B}(x)$ is another independent Brownian motion
and $\equiv$ indicates that the random variables on both sides have the identical
distribution, though they are not equal. Using this property and after a few steps of
elementary algebra, one can rewrite Eq. (\ref{profile}) as,
\begin{equation}
Q_0(x)=1-P(x) = { {Z_{\mu}(\infty)}\over {Z_{\mu}(\infty) + {\tilde Z_{-\mu}}(x)} },
\label{profile1}
\end{equation}
where ${\tilde Z_{-\mu}}(x)=\int_0^x e^{-2\mu x' + 2 {\tilde B}(x')}dx'$. Interestingly, 
exactly  
the same ratio as in Eq. (\ref{profile1}) has appeared earlier in other contexts and its average 
(over disorder) is known exactly\cite{KT1,CMY}. Using these known results and setting $\nu=\mu/\sigma$, we 
get exactly the same asymptotic (large $x$) persistence profile as in Eq. (\ref{profilef}) which
was obtained in Section III by a completely different method. 

\section{Estimation of the energy change due to the second order term in perturbation theory for $\mu<0$}

In this appendix we consider the case $\mu<0$ and provide an estimate of 
second order contribution 
$\Delta E_2$ to the ground state energy $E= 0 + \Delta E_1 +\Delta E_2$ 
of the full Hamiltonian ${\hat H}={\hat H_0}+{\hat 
H_1}$ given by Eq. (\ref{hamil}), treating ${\hat H_1}=ae^{\phi}$ (setting $\lambda=1$) 
as a perturbation to the unperturbed Hamiltonian 
${\hat H_0}=-{1\over {2}}{
{\partial^2}\over {\partial \phi^2}}+ V_{Q}(\phi)$. The quantum potential $V_Q(\phi)$ is given
explicitly by
\begin{equation}
V_Q(\phi)={a\over {2}}\left[a e^{2\phi} + (\nu'-1)e^{\phi} +
4a\left({ {\nu'^2}\over {16a^2}}-p\right)-2p(\nu'+1)e^{-\phi}+4ap^2e^{-2\phi}\right].
\label{qpot1}
\end{equation}
In writing the explicit form of the potential we have substituted the expression of $b(\phi)$ from Eq.
(\ref{bphi}) in Eq. (\ref{hamil}) and used the definition $a=1/{4\sigma}$.
Let us recall that the ground state energy of the unperturbed Hamiltonian $\hat H_0$ is $0$
and the ground state wavefunction is $\psi_0(\phi)$ given explicitly by Eq. (\ref{psi01})
is section-IV. The first order contribution $\Delta E_1= <\psi_0|{\hat H_1}\psi_0>$
was already evaluated exactly in Eq. (\ref{e0}) and was to shown to scale as $\sim p^{\alpha}$
as $p\to 0$ with $\alpha=1$ for $\nu'>1$ and $\alpha=\nu'$ for $\nu'<1$. The goal of this appendix is
to show that the second order contribution $\Delta E_2$ is neglible compared to $\Delta E_1$
as $p\to 0$.   

Note that for $p>0$, it is clear from Eq. (\ref{qpot1}) that 
$V_Q(\phi)\to \infty$ as $\phi\to \pm \infty$. This indicates that for $p>0$ the spectrum of
${\hat H_0}$ is discrete and consists of bound states only.  
Let $\psi_n$'s denote the discrete energy eigenfunctions of 
${\hat H_0}$ with corresponding eigenvalues denoted by $e_n$. The second order contribution 
to the ground state $\Delta E_2$ then follows
from the standard quantum mechanics,  
\begin{equation}
\Delta E_2 = \sum_{n>0}{ {{|<\psi_0|{\hat H_1}|\psi_n>|}^2}\over {e_0-e_n} }=-\sum_{n> 0}
{ {{|<\psi_0|{\hat H_1}|\psi_n>|}^2}\over {e_n}},
\label{de21}
\end{equation}
where we have used the fact that $e_0=0$ as discussed earlier. Using the fact  $e_1<e_2 <e_3 \dots$ in
Eq. (\ref{de21}), one
can immediately obtain an upper bound to $-\Delta E_2$,
\begin{equation}
-\Delta E_2 \le  {1\over {e_1}} \sum_{n>0} |<\psi_0|{\hat H_1}|\psi_n>|^2.
\label{de2bound1}
\end{equation}
By adding and subtracting the $n=0$ term to the sum on the right hand side of the above inequality
and using the completeness of eigenfunctions we get
\begin{equation}
-\Delta E_2 \le {1\over {e_1}}\left[<\psi_0|{\hat H_1}^2|\psi_0>-|<\psi_0|{\hat H_1}|\psi_0>|^2\right].
\label{de2bound2}    
\end{equation}

The quantity $S=<\psi_0|{\hat H_1}^2|\psi_0>-|<\psi_0|{\hat H_1}|\psi_0>|^2$ inside the parenthesis on the 
right hand side of the inequality (\ref{de2bound2})
can be evaluated exactly.  In general, for any $m$, one can express the matrix element
$<\psi_0|{\hat H_1}^m|\psi_0>= a^m \int_{-\infty}^{\infty}\psi_0^2(\phi)e^{m\phi}d\phi$
in the $\phi$ basis by using ${\hat H_1}=ae^{\phi}$. We then substitute the explicit 
expression of 
$\psi_0(\phi)$ from Eq. (\ref{psi01}). The resulting integral can be performed exactly using the 
identity\cite{GR}
\begin{equation}
\int_0^{\infty} x^{\nu-1}e^{-\gamma x -\beta/x}dx= 2 \left({\beta\over 
{\gamma}}\right)^{\nu/2}K_{\nu}\left(2\sqrt(\beta\gamma \right).
\label{iden}
\end{equation}
In fact, throughout this paper, we have heavily used this identity. We then get
$<\psi_0|{\hat H_1}^m|\psi_0>= a^m (2p)^{m/2} 
K_{m-\nu'}\left(\sqrt{2p}/\sigma\right)/K_{\nu'}\left(\sqrt{2p}/\sigma\right)$. 
The expression of $S$ requires the results for $m=2$ and $m=1$ and we get
\begin{equation}
S(p) = 2a^2p\left[ { {K_{2-\nu'}\left(\sqrt{2p}/\sigma\right)}\over{K_{\nu'}\left(\sqrt{2p}/\sigma\right)}} 
-{ {K_{1-\nu'}^2 \left(\sqrt{2p}/\sigma\right)}\over {K_\nu'^2\left(\sqrt{2p}/\sigma\right)} }\right].
\label{snum}
\end{equation}
Using the expansion of Bessel functions for small arguments\cite{GR}, it is easy to show that 
in the limit $p\to 0$, $S(p)\sim p^2$ for $\nu'>2$, $S(p)\to -p^2\log p $ for $\nu'=2$ and
$S(p)\sim p^{\nu'}$ for $\nu'<2$. 

Having established the behavior of $S$ for small $p$, we now need to estimate the gap $e_1$
(the energy of the first excited state of ${\hat H_0}$) for small $p$  
on the right hand side of the inequality $-\Delta E_2 \le S(p)/e_1$ in Eq. (\ref{de2bound2}). 
To estimate the gap,
we examine the quantum potential in Eq. (\ref{qpot1}). It is convenient first
to make a change of variable
$z=e^{\phi}$ so that $0\le z\le \infty$. In this new variable the quantum potential in Eq. (\ref{qpot1})
reads,
\begin{equation}
V_Q(z) = {a\over {2}}\left[a z^2 + (\nu'-1)z + 4a\left({{\nu'^2}\over {16 a^2}}-p\right)-{{2p(\nu'+1)}\over 
{z}}+{{4a p^2}\over 
{z^2}}\right].
\label{qpot2}
\end{equation}    
The shape of this potential is shown in Fig. 2. Note that the potential $V_Q(z)$ has 
a minimum at $z=z_0$, determined from the equation $dV_Q(z)/dz=0$ which gives,
$2a z_0^4 +(\nu'-1)z_0^3 + 2p(\nu'+1)z_0 -8ap^2=0$. In the limit $p\to 0$, the 
only real root of this equation is at $z_0\approx 4ap/(\nu'+1)$. The value of the potential 
at this minimum, $V_Q(z_0)\to -(2\nu'+1)/8 +O(p)$ as $p\to 0$. We also need to estimate the typical
width of the potential $W(e)$ at an energy $e$ for small $p$ (see Fig. 2). 
The points $z_{\pm}(e)$ where $V_Q(z)=e$ can be easily estimated for small
$p$ since in this limit one just needs to solve a quadratic equation and we get, 
\begin{equation}
z_{\pm}(e)\approx {{4a}\over {\left[\nu'+1\mp \sqrt{2\nu'+1+2e}\right]} }p +O(p^2).  
\label{roots}
\end{equation}
Hence the typical width of the potential at energy $e$ scales as $W(e)\approx z_+(e)-z_{-}(e)\sim p$
in the $p\to 0$ limit at any finite level $e$.
For small $p$, one can approximate the potential $V_Q(z)$ around its mimimum $z=z_0$ by a 
harmonic oscillator potential, $V_q(z)\approx -(2\nu'+1)/8 + \omega^2 z^2 /2$ where the frequency
$\omega$ is estimated from the typical width , i.e., $ \omega^2 W^2/2 \sim O(1)$.
Since $W(e)\sim p$, we find that the frequency scales as $\omega\sim 1/p$ as $p\to 0$.
One knows that the gap between the first excited state and the ground state in a harmonic potential scales 
as $e_1\sim \omega$.
Thus we estimate that the energy of the first excited state scales as $e_1\sim \omega\sim 1/p$
in the limit $p\to 0$.    

Substituting this estimate of the gap in the inequality, $-\Delta E_2\le S(p)/e_1$ and using 
the small $p$ estimates of $S(p)$ derived earlier, we find that as $p\to 0$, 
$-\Delta E_2\le p^3$ for $\nu'>2$, $-\Delta E_2\le -p^3\log p$ for $\nu'=2$ and
$-\Delta E_2\le p^{\nu'+1}$ for $\nu'<1$. Comparing these results with the 
first order contribution where $\Delta E_1 \sim p$ for $\nu'>1$ and
$\Delta E_1 \sim p^{\nu'}$ for $\nu'<1$, we conclude that the second order
contribution is negligible compared to the first order term in the 
limit $p\to 0$.

\newpage
\begin{figure}
\begin{center}
\leavevmode
\psfig{figure=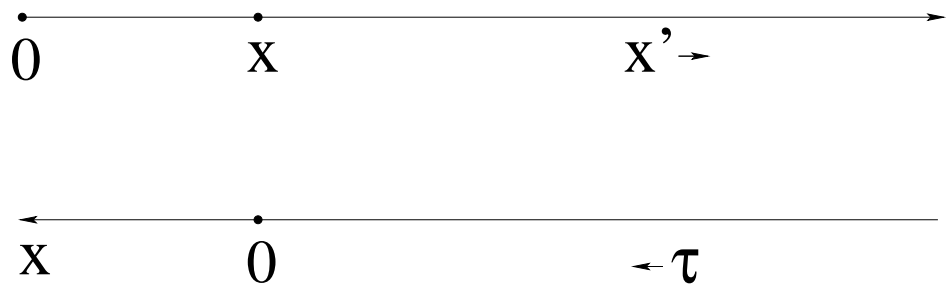,width=14cm,angle=0}
\caption{The change of variable $\tau=x-x'$ for fixed $x$. The new variable
$\tau$ increases from $\tau=-\infty$ (when $x'=\infty$) to $\tau=x$ (when $x'=0$)
through the point $\tau=0$ (when $x'=x$).}
\end{center}
\end{figure}                                  

\begin{figure}
\begin{center}
\leavevmode
\psfig{figure=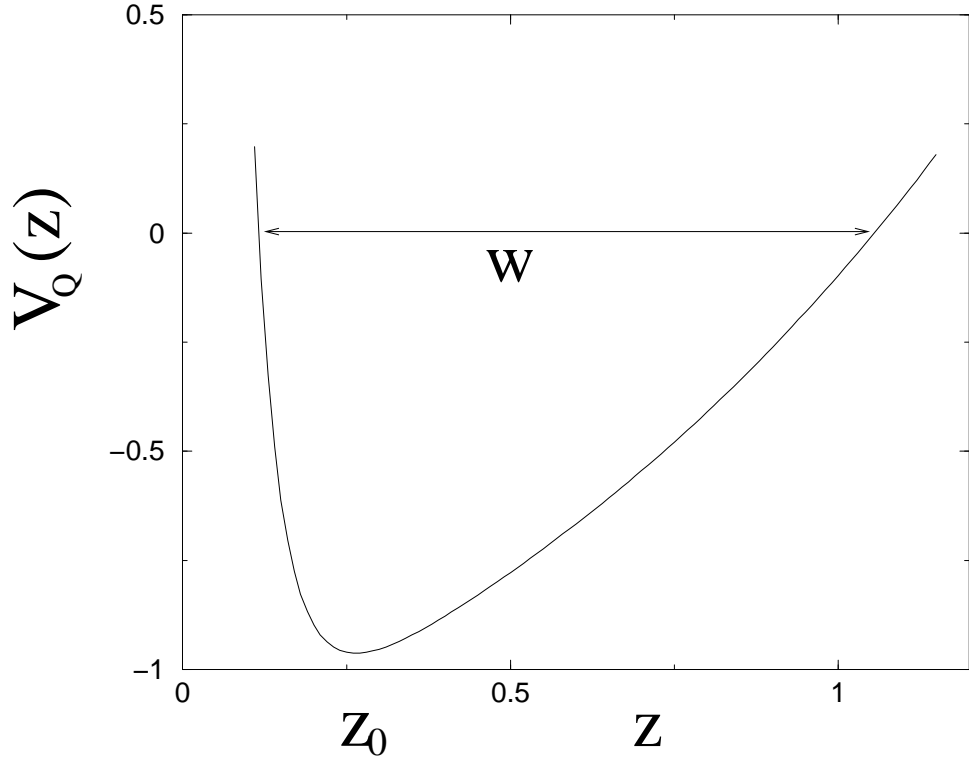,width=14cm,angle=0}
\caption{The shape of the potential $V_Q(z)$ in Eq. (\ref{qpot2}) is shown for
parameter values: $a=1$, $\nu'=0.5$ and $p=0.1$. The potential has a minimum
at $z=z_0\sim p$ and a typical width $W\sim p$ in the limit $p\to 0$.}
\end{center}
\end{figure}

\end{document}